


\font\bigbold=cmbx12
\font\ninerm=cmr9
\font\eightrm=cmr8
\font\sixrm=cmr6
\font\fiverm=cmr5
\font\ninebf=cmbx9
\font\eightbf=cmbx8
\font\sixbf=cmbx6
\font\fivebf=cmbx5
\font\ninei=cmmi9  \skewchar\ninei='177
\font\eighti=cmmi8  \skewchar\eighti='177
\font\sixi=cmmi6    \skewchar\sixi='177
\font\fivei=cmmi5
\font\ninesy=cmsy9 \skewchar\ninesy='60
\font\eightsy=cmsy8 \skewchar\eightsy='60
\font\sixsy=cmsy6   \skewchar\sixsy='60
\font\fivesy=cmsy5
\font\nineit=cmti9
\font\eightit=cmti8
\font\ninesl=cmsl9
\font\eightsl=cmsl8
\font\ninett=cmtt9
\font\eighttt=cmtt8
\font\tenfrak=eufm10
\font\ninefrak=eufm9
\font\eightfrak=eufm8
\font\sevenfrak=eufm7
\font\fivefrak=eufm5
\font\tenbb=msbm10
\font\ninebb=msbm9
\font\eightbb=msbm8
\font\sevenbb=msbm7
\font\fivebb=msbm5
\font\tensmc=cmcsc10


\newfam\bbfam
\textfont\bbfam=\tenbb
\scriptfont\bbfam=\sevenbb
\scriptscriptfont\bbfam=\fivebb
\def\Bbb{\fam\bbfam}

\newfam\frakfam
\textfont\frakfam=\tenfrak
\scriptfont\frakfam=\sevenfrak
\scriptscriptfont\frakfam=\fivefrak
\def\frak{\fam\frakfam}

\def\smc{\tensmc}


\def\eightpoint{%
\textfont0=\eightrm   \scriptfont0=\sixrm
\scriptscriptfont0=\fiverm  \def\rm{\fam0\eightrm}%
\textfont1=\eighti   \scriptfont1=\sixi
\scriptscriptfont1=\fivei  \def\oldstyle{\fam1\eighti}%
\textfont2=\eightsy   \scriptfont2=\sixsy
\scriptscriptfont2=\fivesy
\textfont\itfam=\eightit  \def\it{\fam\itfam\eightit}%
\textfont\slfam=\eightsl  \def\sl{\fam\slfam\eightsl}%
\textfont\ttfam=\eighttt  \def\tt{\fam\ttfam\eighttt}%
\textfont\frakfam=\eightfrak \def\frak{\fam\frakfam\eightfrak}%
\textfont\bbfam=\eightbb  \def\Bbb{\fam\bbfam\eightbb}%
\textfont\bffam=\eightbf   \scriptfont\bffam=\sixbf
\scriptscriptfont\bffam=\fivebf  \def\bf{\fam\bffam\eightbf}%
\abovedisplayskip=9pt plus 2pt minus 6pt
\belowdisplayskip=\abovedisplayskip
\abovedisplayshortskip=0pt plus 2pt
\belowdisplayshortskip=5pt plus2pt minus 3pt
\smallskipamount=2pt plus 1pt minus 1pt
\medskipamount=4pt plus 2pt minus 2pt
\bigskipamount=9pt plus4pt minus 4pt
\setbox\strutbox=\hbox{\vrule height 7pt depth 2pt width 0pt}%
\normalbaselineskip=9pt \normalbaselines
\rm}


\def\ninepoint{%
\textfont0=\ninerm   \scriptfont0=\sixrm
\scriptscriptfont0=\fiverm  \def\rm{\fam0\ninerm}%
\textfont1=\ninei   \scriptfont1=\sixi
\scriptscriptfont1=\fivei  \def\oldstyle{\fam1\ninei}%
\textfont2=\ninesy   \scriptfont2=\sixsy
\scriptscriptfont2=\fivesy
\textfont\itfam=\nineit  \def\it{\fam\itfam\nineit}%
\textfont\slfam=\ninesl  \def\sl{\fam\slfam\ninesl}%
\textfont\ttfam=\ninett  \def\tt{\fam\ttfam\ninett}%
\textfont\frakfam=\ninefrak \def\frak{\fam\frakfam\ninefrak}%
\textfont\bbfam=\ninebb  \def\Bbb{\fam\bbfam\ninebb}%
\textfont\bffam=\ninebf   \scriptfont\bffam=\sixbf
\scriptscriptfont\bffam=\fivebf  \def\bf{\fam\bffam\ninebf}%
\abovedisplayskip=10pt plus 2pt minus 6pt
\belowdisplayskip=\abovedisplayskip
\abovedisplayshortskip=0pt plus 2pt
\belowdisplayshortskip=5pt plus2pt minus 3pt
\smallskipamount=2pt plus 1pt minus 1pt
\medskipamount=4pt plus 2pt minus 2pt
\bigskipamount=10pt plus4pt minus 4pt
\setbox\strutbox=\hbox{\vrule height 7pt depth 2pt width 0pt}%
\normalbaselineskip=10pt \normalbaselines
\rm}


\def\pagewidth#1{\hsize= #1}
\def\pageheight#1{\vsize= #1}
\def\hcorrection#1{\advance\hoffset by #1}
\def\vcorrection#1{\advance\voffset by #1}

\newif\iftitlepage   \titlepagetrue               
\newtoks\titlepagefoot     \titlepagefoot={\hfil} 
\newtoks\otherpagesfoot    \otherpagesfoot={\hfil\tenrm\folio\hfil}
\footline={\iftitlepage\the\titlepagefoot\global\titlepagefalse
           \else\the\otherpagesfoot\fi}

\font\extra=cmss10 scaled \magstep0
\setbox1 = \hbox{{{\extra R}}}
\setbox2 = \hbox{{{\extra I}}}
\setbox3 = \hbox{{{\extra C}}}
\setbox4 = \hbox{{{\extra Z}}}
\setbox5 = \hbox{{{\extra N}}}

\def\RRR{{{\extra R}}\hskip-\wd1\hskip2.0 
   true pt{{\extra I}}\hskip-\wd2
\hskip-2.0 true pt\hskip\wd1}
\def\Real{\hbox{{\extra\RRR}}}    

\def\CCC{{{\extra C}}\hskip-\wd3\hskip 2.5 true pt{{\extra I}}
\hskip-\wd2\hskip-2.5 true pt\hskip\wd3}
\def\Complex{\hbox{{\extra\CCC}}\!\!}   







\def\R{{\Real}}
\def\C{{\Complex}}


\def\bfA{A\hskip-7.35pt A}
\def\bfB{B\hskip-7.85pt B}
\def\bfE{E\hskip-7.85pt E}
\def\bfS{S\hskip-7.1pt S}

\def\bfa{\hbox{$a\!\!\!a$}}
\def\bfb{b\hskip-4pt b}
\def\bfc{c\hskip-4pt c}

\def\bff{f\hskip-5.67pt f}
\def\bfg{g\hskip-5.5pt g}
\def\bfh{h\hskip-5.4pt h}

\def\bfp{p\hskip-5.5pt p}

\def\bfr{r\hskip-5.1pt r}
\def\bfs{\hbox{$s\hskip-5.1pt s$}}

\def\bfu{\hbox{$u\hskip-5.5pt u$}}
\def\bfv{\hbox{$v\hskip-5.65pt v$}}
\def\bfw{\hbox{$w\hskip-7.18pt w$}}
\def\bfx{x\hskip-5.4pt x}
\def\bfy{y\!\!\!y}


\def\exbfa{a\hskip-4.2pt a}
\def\exbfb{b\hskip-3.9pt b}
\def\exbfc{c\hskip-4pt c}
\def\exbfh{h\hskip-4.9pt h}
\def\exbfr{r\hskip-4.15pt r}
\def\exbfu{u\hskip-4.9pt u}
\def\exbfx{x\hskip-4.3pt x}
\def\exbfrho{\rho\hskip-4.3pt \rho}


\def\bfchi{\chi\hskip-6.62pt \chi}
\def\bfrho{\rho\hskip-4.9pt \rho}
\def\bfeta{\eta\hskip-5pt \eta}

\def\bfzero{0\hskip-5.5pt 0}


\def\frac#1#2{{#1\over#2}}

\def\({\left(}
\def\){\right)}
\def\<{\langle}
\def\>{\rangle}

\def\pmb#1{\setbox0=\hbox{$#1$}%
   \kern-.025em\copy0\kern-\wd0
   \kern.05em\copy0\kern-\wd0
   \kern-.025em\raise.0433em\box0 }


\def\abstract#1{{\parindent=30pt\narrower\noindent\ninepoint\openup
2pt #1\par}}


\newcount\notenumber\notenumber=1
\def\note#1
{\unskip\footnote{$^{\the\notenumber}$}
{\eightpoint\openup 1pt #1}
\global\advance\notenumber by 1}


\global\newcount\secno \global\secno=0
\global\newcount\subsecno \global\subsecno=0
\global\newcount\meqno \global\meqno=1
\global\newcount\appno \global\appno=0
\newwrite\eqmac
\def\romappno{\ifcase\appno\or A\or B\or C\or D\or E\or F\or G\or H
\or I\or J\or K\or L\or M\or N\or O\or P\or Q\or R\or S\or T\or U\or
V\or W\or X\or Y\or Z\fi}
\def\eqn#1{
        \ifnum\secno>0
          \ifnum\subsecno<1
            \eqno(\the\secno.\the\meqno)\xdef#1{\the\secno.\the\meqno}
          \else
            \eqno(\the\secno.\the\subsecno.\the\meqno)
\xdef#1{\the\secno.\the\subsecno.\the\meqno}
          \fi
          \else\ifnum\appno>0
            \eqno({\rm\romappno}.\the\meqno)
             \xdef#1{{\rm\romappno}.\the\meqno}
          \else
            \eqno(\the\meqno)\xdef#1{\the\meqno}
          \fi
        \fi
\global\advance\meqno by1 }


\global\newcount\refno
\global\refno=1 \newwrite\reffile
\newwrite\refmac
\newlinechar=`\^^J
\def\ref#1#2{\the\refno\nref#1{#2}}
\def\nref#1#2{\xdef#1{\the\refno}
\ifnum\refno=1\immediate\openout\reffile=refs.tmp\fi
\immediate\write\reffile{
     \noexpand\item{[\noexpand#1]\ }#2\noexpand\nobreak.}
     \immediate\write\refmac{\def\noexpand#1{\the\refno}}
   \global\advance\refno by1}
\def\semi{;\hfil\noexpand\break ^^J}
\def\nl{\hfil\noexpand\break ^^J}
\def\refn#1#2{\nref#1{#2}}

\def\ann#1#2#3{{\it Ann. Phys.} {\bf {#1}} (19{#2}) #3}
\def\cmp#1#2#3{{\it Commun. Math. Phys.} {\bf {#1}} (19{#2}) #3}

\def\jmp#1#2#3{{\it J. Math. Phys.} {\bf {#1}} (19{#2}) #3}
\def\jpA#1#2#3{{\it J.  Phys.} {\bf A{#1}} (19{#2}) #3} 
 
\def\ijtp#1#2#3{{\it Int.  J.  Theor.  Phys.} {\bf {#1}} (19{#2}) #3}

\def\np#1#2#3{{\it Nucl.  Phys.} {\bf B{#1}} (19{#2}) #3} 
 
\def\plA#1#2#3{{\it Phys.  Lett.} {\bf {#1}A} (19{#2}) #3}

\def\prD#1#2#3{{\it Phys.  Rev.} {\bf D{#1}} (19{#2}) #3}

\def\prp#1#2#3{{\it Phys.  Rep.} {\bf {#1}C} (19{#2}) #3}


{

\refn\MaF
{For a review, see, {\it e.g.}, 
V.P. Maslov and M.V. Fedoriuk, 
\lq\lq Semiclassical Approximation in Quantum 
Mechanics\rq\rq, 
D. Reidel Publ., London, 1981}

\refn\KO
{For a recent review, see 
Y.A. Kravtsov and Y.I. Orlov,
\lq\lq Caustics, Catastrophes and Wave Fields\rq\rq
(Second Edition), 
Springer, Berlin, 1999}

\refn\Berry
{M.V. Berry, 
{\sl Adv. Phys.} {\bf 25} (1976) 1}

\refn\Schulman
{L.S. Schulman, \lq\lq Techniques and Applications of
Path Integration\rq\rq, John Wiley \& Sons, New York, 1981}

\refn\S
{L.S. Schulman, in \lq\lq Functional Integration
and its Applications\rq\rq, A.M. Arthurs, ed., 
Clarendon Press, Oxford, 1975}

\refn\DV
{G. Dangelmayr and W. Veit,
\ann{118}{79}{108}}

\refn\DMMN
{C. DeWitt-Morette, A. Maheshwari and B. Nelson, 
\prp{50}{79}{256}}

\refn\LS
{S. Levit and U. Smilansky,
{\sl Proc. Amer. Math. Soc.} {\bf 21} (1977) 299;
\ann{103}{77}{198}}

\refn\DM
{C. DeWitt-Morette, \ann{97}{76}{367}}

\refn\Souriau
{J.-M. Souriau, in \lq\lq Group Theoretical Methods
in Physics\rq\rq, A. Janner, T. Janssen and M. Boon, eds., 
Lecture Notes in Physics, {\bf 50}, Springer-Verlag, Berlin, 1976}

\refn\Horv
{P.A. Horv\'{a}thy, \ijtp{13}{79}{245}}

\refn\C
{B.K. Cheng, \plA{101}{84}{464}}

\refn\FC
{R. Ferreira and B.K. Cheng, \jpA{18}{85}{L1127}}

\refn\HMTT
{K. Horie, H. Miyazaki, I. Tsutsui and S. Tanimura,
\ann{273}{99}{267}}

\refn\GY
{I.M. Gel'fand and A.M. Yaglom, \jmp{1}{60}{48}}

\refn\HMTTlet
{K. Horie, H. Miyazaki, I. Tsutsui and S. Tanimura,
\plA{253}{99}{259}}

\refn\Witten
{E. Witten, \np{149}{79}{285}}

\refn\Jevicki
{A. Jevicki, \prD{20}{79}{3331}}

\refn\CH
{R. Courant and D. Hilbert, 
\lq\lq Methods of Mathematical Physics\rq\rq, 
Interscience Publishers, New York, 1953}

\refn\MF
{P.M. Morse and H. Feshbach,
\lq\lq Methods of Theoretical Physics\rq\rq, 
McGraw-Hill Book Company, New York, 1953}

\refn\Milnor
{J. Milnor, \lq\lq Morse Theory\rq\rq, 
Princeton University Press, Princeton, 1963}

\refn\FH
{R.P. Feynman and A.R. Hibbs, 
\lq\lq Quantum Mechanics and Path Integrals\rq\rq, 
McGraw-Hill, New York, 1965}

\refn\H
{G.A. Hagedorn, \cmp{71}{80}{77}}





}

\def\ve{\vfill\eject}




\pageheight{23cm}
\pagewidth{14.8cm}
\hcorrection{0mm}
\magnification= \magstep1
\def\bsk{%
\baselineskip= 16.8pt plus 1pt minus 1pt}
\parskip=5pt plus 1pt minus 1pt
\tolerance 6000


\null


{
\leftskip=100mm
\hfill\break
KEK Preprint 99-18
\hfill\break
\par}

\smallskip
\vfill
{\baselineskip=18pt

\centerline{\bigbold 
Quantum Caustics for Systems with Quadratic Lagrangians}
\centerline{\bigbold in Multi-Dimensions}

\vskip 40pt

\centerline{
\smc 
Kenichi Horie,
\quad 
Hitoshi Miyazaki
\quad {\rm and} \quad 
Izumi Tsutsui\note
{E-mail:\quad izumi.tsutsui@kek.jp}
}

\vskip 5pt

{
\baselineskip=13pt
\centerline{\it 
Institute of Particle and Nuclear Studies}
\centerline{\it 
High Energy Accelerator Research Organization (KEK),
Tanashi Branch}
\centerline{\it Tokyo 188-8501, Japan}
}

\vskip 70pt

\abstract{%
{\bf Abstract.}\quad
We study quantum 
caustics 
({\it i.e.}, the quantum analogue of 
the classical singularity in 
the Dirichlet boundary problem)
in $d$-dimensional systems
with quadratic Lagrangians of the form
$L = {1\over 2} P_{ij}(t)\, \dot x^i \dot x^j
+ Q_{ij}(t)\, x^i \dot x^j + {1\over 2} R_{ij}(t) \, x^i x^j
+ S_i(t)\, x^i$.  Based on Schulman's procedure in the path-integral
we derive the transition amplitude on caustics
in a closed form for generic multiplicity $f$,
and thereby complete 
the previous analysis carried out
for the maximal multiplicity case ($f = d$).
The unitarity relation, together with the initial condition, 
fulfilled by the amplitude
is found to be a key ingredient for determining 
the amplitude, which reduces 
to the well-known
expression with Van-Vleck determinant for the non-caustics
case ($f = 0$).
Multiplicity dependence of the caustics phenomena
is illustrated by examples of a particle interacting
with external electromagnetic fields.
}

\bigskip
{\ninepoint
PACS codes: 02.30.Wd; 03.65.-w; 03.65.Sq \hfill\break 
\indent
{Keywords: Caustics, Semiclassical Approximations, Path-integral}
}


\pageheight{23cm}
\pagewidth{15.7cm}
\hcorrection{-1mm}
\magnification= \magstep1
\def\bsk{%
\baselineskip= 16.5pt plus 1pt minus 1pt}
\parskip=5pt plus 1pt minus 1pt
\tolerance 8000
\bsk



\ve

\secno=1 \meqno=1 


\centerline{\bf 1. Introduction}
\medskip

The semiclassical approximation of quantum mechanics [\MaF]
is a powerful means to investigate various --- often
non-perturbative --- aspects of quantization in reference
to classical mechanics.  In the path-integral framework, 
it amounts
to keeping only up to quadratic terms in the
Lagrangian, and as such the core of
the semiclassical approximation resides in the evaluation
of the integral for quadratic systems.  
The path-integral then becomes Gaussian 
allowing for a closed form for the
transition amplitude in terms of the familiar Van Vleck determinant.
That is, in $d$-dimensions, 
the amplitude for the transition between $\bfa$, 
$\bfb \in \R^d$ during the time interval $[0, T]$ 
reads
$$
K (\bfb,T;\bfa,0) = 
\sqrt{
\det\left( {i\over{2\pi \hbar}}
{{\partial^2 I[\bar{\bfx}]}\over
{\partial a^i \partial b^j}}  
\right)} 
\,e^{ {i\over\hbar}I[\bar{\exbfx}] }\ .
\eqn\vvformula
$$
Here $I[\bar{\bfx}] = \int^T_0 dt\, L$ 
is the action for the quadratic
part of the Lagrangian along
the classical path $\bar{\bfx}(t)$ which satisfies the Dirichlet 
boundary conditions corresponding to the transition.
The point to be noted is that 
the transition kernel $K (\bfb,T;\bfa,0)$ in (\vvformula)
develops singularities if there exists no (or more than one)
classical paths which meet the Dirichlet boundary conditions, 
and accordingly there arise 
physical phenomena characteristic to the singularities
known as {\it caustics} in geometrical optics [\KO]. 

Historically, the semiclassical analysis of 
caustics phenomena was conducted intensively
in late seventies in association with  
the catastrophe theory 
(see, {\it e.g.}, [\Berry, \Schulman] and references therein).  
There, one takes into account 
cubic terms to avoid the singularities (and hence the caustics 
no longer exist in the strict sense),
which is in fact an appropriate procedure 
for most realistic physical systems.  The path-integral
is then approximated by the 
\lq generalized Airy integral\rq~governed 
by catastrophe polynomials [\S, \DV, \DMMN]. 
On the other hand, for endpoints {\it beyond caustics}, namely
when the endpoints go beyond singular (conjugate) points, 
the analysis for pure quadratic systems
was carried out in evaluating the correction in phase factor 
[\LS] as well as in providing
a more general basis for the semiclassical expansion in the 
path-integral [\DM].  The latter considered also endpoints 
{\it on caustics}, but only for the extreme 
case when the caustics occur maximally, 
that is, when the multiplicity $f$ 
of the caustics coincides with the
dimension $d$.  Specific examples on caustics
have been studied independently 
in [\Souriau, \Horv] (see also [\C, \FC]).

The aim of this paper is to present a complete
analysis of quantum caustics for quadratic systems
by deriving the transition amplitude in a closed form 
under generic multiplicity $f$ 
on caustics in $d$-dimensions,
extending that of [\DM] and of our own [\HMTT].
We shall find that, in both classical and quantum regimes, 
caustics have a rich
structure characterized by the multiplicity, and this
will be demonstrated by two examples furnished later.
Schulman's procedure [\S, \Schulman] will be adopted for 
evaluating the kernel on caustics, but the key ingredient
to get the closed form turns out to be the unitarity
relation and the initial condition 
satisfied by the kernel.  This we find is 
amusing, as the set of these requirements alone is 
sufficient to get the kernel formula (\vvformula) for 
regular (non-caustics) cases without recourse to any involved
measures.   

This paper is organized as follows.  After the Introduction,
in Section 2 we provide
a general argument for caustics both in
classical and quantum mechanics.  Using the unitarity relation 
as a key ingredient, the transition amplitude
is derived explicitly for 
the generic case of multiplicity.  For illustration, 
in Section 3 we consider a particle moving under external
electric and magnetic fields exhibiting caustics phenomena 
with different multiplicities.
Section 4 is devoted to our conclusion.

\ve

\secno=2 \meqno=1 


\centerline{\bf 2. Caustics in Classical and Quantum Mechanics}
\medskip

In this section we present a general theory of caustics for 
quadratic systems in $d$-dimensions
in both classical mechanics and quantum mechanics.
We first provide a framework for classical caustics which are
classified by the multiplicity $f$ given by
the co-dimension of the surface formed by
the set of focal points.  
We then derive the transition kernel for 
generic $f$ based on  
the unitarity relation and the initial condition
satisfied by the kernel.
The result reduces to the standard formula (\vvformula) for 
the nonsingular ($f = 0$) case 
and to the one previously obtained for 
the maximally singular ($f = d$) case.

\bigskip
\noindent
{\bf 2.1. Classical caustics}

The systems we are interested in are 
those described by Lagrangians
in $d$-dimensions which are at most quadratic:
$$
L = {1\over 2} P_{ij}(t)\, \dot x^i \dot x^j
+ Q_{ij}(t)\, x^i \dot x^j + {1\over 2} R_{ij}(t) \, x^i x^j
+ S_i(t)\, x^i\ .
\eqn\lagrangian
$$
The coefficient functions $P_{ij}(t)$, $Q_{ij}(t)$, 
$R_{ij}(t)$, $S_i(t)$
with $i,\, j = 1, 2, \ldots, d$ 
are smooth functions of time $t$, and it is understood that 
repeated indices are summed over unless otherwise stated.  
We take $P_{ij}(t)$ and $R_{ij}(t)$ symmetric 
as a matrix, $P^T(t) = P(t)$, $R^T(t) = R(t)$, 
and assume that $P(t)$ is positive-definite,
{\it i.e.,} all of its minors are positive for any 
$t \in [0, T]$.
The equations of motion derived from the action read
$$
\Lambda\, \bfx + \bfS = \bfzero\ ,
\eqn\em
$$
where we have used the matrix-valued operator,
$$
\Lambda
 := - {{d}\over{dt}}\left(P(t) {{d}\over{dt}} 
    + Q^T(t)\right) + \left(Q(t) {d\over{dt}}  + R(t)\right)\ .
\eqn\operator
$$
We note that the operator $\Lambda$ 
is self-adjoint in the space of functions
which vanish at the time boundary $t = 0$ and $T$.

We shall consider the Dirichlet problem associated with
the equation (\em), that is, we look for the solution
$\bar{\bfx}(t)$ of (\em) satisfying the boundary conditions,
$$
\bar{\bfx}(0) = \bfa, \qquad  
\bar{\bfx}(T) = \bfb, \qquad
\eqn\bc
$$
to some given vectors $\bfa$, $\bfb \in \R^d$.  
The usual procedure for this is to
choose first two independent sets of solutions,
$\{ \bfv_k(t) \}$ and $\{ \bfu_k(t) \}$ for 
$k = 1, 2, \ldots, d$,
which obey the homogeneous (Jacobi) equation,
$$
\Lambda\, \bfv_k = \bfzero, \qquad
\Lambda\, \bfu_k = \bfzero, \qquad
k = 1, 2, \ldots, d\ .
\eqn\jacobieq
$$
The set $\{ \bfv_k \}$ is specified by requiring that
$$
v^i_k(0) = \delta_{ik}, \qquad \dot v^i_k(0) = 0,
\qquad i, k = 1, 2, \ldots, d\ .
\eqn\vsol
$$
On the other hand, for $\{ u_k \}$ 
we impose only the conditions,
$$
u^i_k(0) = 0, \qquad
i, k = 1, 2, \ldots, d\ ,
\eqn\usol
$$
{}for the moment and leave $\dot u^i_k(0)$ undetermined.  
These solutions form a complete set of solutions for the
homogeneous equation.\note{%
Note that our boundary conditions, (\vsol) and
(\usol), differ slightly from those used in Ref.[\DM].
}
Later, we shall impose other conditions to specify the set 
uniquely.  

We also choose a special solution $\bfs(t)$ of the full
equations of motion (\em), $\Lambda\, \bfs + \bfS = 0$, 
obeying the initial condition $s^i(0) = 0$.  
(For our present purpose we do not
need to specify the initial velocities $\dot s^j(0)$.)
Then, in terms of $2d$ constants, 
$A^k$, $B^k$ for $k = 1, 2, \ldots, d$,  
the general solution of the equations of motion (\em)
is given by
$$
\bar{\bfx}(t) = A^k\, \bfu_k(t) + B^k\, \bfv_k(t) + \bfs(t)\ .
\eqn\general
$$
{}For convenience, we
introduce the matrices $U(t)$ and $V(t)$ from the solutions
by $U_{ik}(t) := u_k^i(t)$ and $V_{ik}(t) := v_k^i(t)$,
respectively, and thereby rewrite the 
general solution (\general) as
$$
\bar{\bfx}(t) = U(t)\, \bfA + V(t)\, \bfB + \bfs(t)\ ,
\eqn\generalb
$$
with $\bfA = (A_1, \ldots, A_d)^T$ and 
$\bfB = (B_1, \ldots, B_d)^T$.
Then the initial condition in (\bc) implies
$\hbox{$\bfB$} = \bfa$ whereas 
the final condition in (\bc) is met by choosing
$$
\bfA = U^{-1}(T)\left(\bfb - V(T)\, \bfa - \bfs(T)\right)\ .
\eqn\choicea
$$
It is thus clear 
that the solution to the Dirichlet problem (\bc)
does not exist if $\det U(T) = 0$, and 
this is the cause of caustics.

To formulate the caustics more precisely, let us define $f$ 
such that $d - f$ 
becomes the rank of the matrix $U(T)$, and choose
the (new) set of solutions $\{ \bfu_k \}$ satisfying
$$
\bfu_k(T) = \bfzero 
\qquad \hbox{for} \quad k = 1, 2, \ldots, f.
\eqn\ucaust
$$
We also demand, for simplicity,
that the non-vanishing  
part of the solutions in the set  be normalized as
$$
u_k^i(T) = \delta_{ik} \qquad \hbox{for} 
\quad i, k = f+1, f+2, \ldots, d\ .
\eqn\unorm
$$
Hence, at $t = T$ the matrix $U(T)$ takes the form,
$$
U(T) = \pmatrix{
0&\ast\cr
0& 1  \cr}.
\eqn\matu
$$
Here the upper left \lq 0' is a $f \times f$ null matrix,
the lower left \lq 0' is a $(d-f) \times f$ null matrix,
\lq 1' is a $(d-f) \times (d-f)$ identity matrix, and 
\lq $\ast$' represents a $f \times (d-f)$ matrix 
consisting of undetermined elements. 
As we shall see shortly, 
the number $f$ gives the co-dimension
of the surface where caustics occur, with $f = d$ 
and $f = 0$ 
being the two extremes.  We call these extreme cases 
\lq full caustics'
and \lq non-caustics', respectively, and use
\lq partial caustics' for other intermediate cases.

Note that
the choice, (\ucaust) and (\unorm), can always be
realized by linear transformations in the
solution space formed by $\{ {\bfu}_k \}$ after some exchanges
of the coordinates, if necessary.
It then follows from the solution (\general) that 
the surface of caustics ({\it i.e.}, 
the set of all endpoints) conjugate to
the initial point $a$ is given by 
$$
\bfx = 
\bfu_k(T) A^k + \bfh(\bfa)\ ,
\eqn\cstsurface
$$
where $A^k$, $k = f+1, \ldots, d$, parameterize the 
$(d - f)$-dimensional caustic surface, and 
we have used 
$$
\bfh(\bfa) := \bfv_k(T) a^k + \bfs(T)\ .
\eqn\ef
$$

Thus, if $f \ne 0$, then for our Dirichlet problem
the best we can do (in our coordinate frame) 
is to find a solution
that meets the conditions ${\bar x}^i(T) = b^i$ for 
$i = f+1, \ldots, d$ by choosing
$$
A^k = A^k(\bfb) := b^k - h^k(\bfa) \qquad \hbox{for} \quad 
k = f+1, f+2, \ldots, d.
\eqn\asol
$$
To specify the solution uniquely, for the remaining
components $k = 1, \ldots, f$ we set $A^k = 0$ for 
simplicity.
With this choice the classical solution turns out to be
$$
{\bar x}^i(t) 
= \Bigl\{v_l^i(t) - \sum_{k = f+1}^d v_l^k(T) u^i_k(t)
  \Bigr\}\, a^l  
+ \sum_{k = f+1}^d u^i_k(t)\, b^k
+ s^i(t) - \sum_{k = f+1}^d u^i_k(t)\, s^k(T)\ ,
\eqn\oursol
$$
whose endpoint,
$$
{\bar x}^i(T) 
= u^i_k(T)\, b^k + 
\left\{ \delta_k^i - u^i_k(T)\right\} h^k(a),
\qquad i = 1, 2, \ldots, d,
\eqn\endp
$$
reduces to $b^i$ for $i = f+1, f+2, \ldots, d$ as required.

An important point to note is that the matrix 
$U(T)$, normalized as (\matu), fulfills 
$(U(T))^2 = U(T)$ and hence it
acts as a projection operator onto the caustic surface
given by (\cstsurface).  Accordingly, the matrix
$U^{\bot}(T) := 1 - U(T)$ projects vectors down to
the complenetary space `orthogonal' to the surface. 
(Strictly speaking, the space is not quite orthogonal
to the surface because $U(T)$ may not be symmetric.)   
In terms of these projection operators
the endpoint of our classical solution (\endp) is
$$
\bar{\bfx}(T) = U(T)\, \bfb + U^{\bot}(T)\, \bfh(\bfa),
\eqn\proj
$$
where now its geometrical meaning is evident (Fig.1).
\topinsert
\vskip 1cm
\let\picnaturalsize=N
\def\picsize{6cm}
\def\picfilename{f1.epsf}
\ifx\nopictures Y\else{\ifx\epsfloaded Y\else\input epsf \fi
\global\let\epsfloaded=Y
\ifx\picnaturalsize N\epsfxsize \picsize\fi
\hskip 2.5cm\epsfbox{\picfilename}}\fi
\vskip -1cm
\abstract{%
{\bf Figure 1.} 
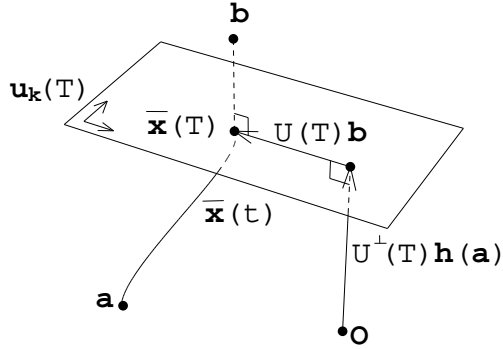
A schematic picture of a caustics plane and 
the decomposition of the endpoint vector $\bar{\bfx}(T)$
by the projection operator.  The vectors 
$U^\bot(T)\exbfh(\bfa)$ and
$\hbox{$\exbfb$} - \bar{\hbox{$\bfx$}}(T)$ become orthogonal to 
$U(T)\exbfb$ when $U(T)$ is symmetric.
}
\endinsert
We here derive some useful identities which will be
important later.
Notice first that, for arbitrary vector-valued functions
$\bff$ and $\bfg$, we have
$$
\int_0^T dt\, (\bff\cdot \Lambda\, \bfg 
- \bfg\cdot \Lambda\, \bff) 
= \left[ \left( \dot{\bff}\cdot P \bfg 
- \dot{\bfg}\cdot P \bff \right) +
  \bff \cdot \left( Q - Q^T\right) \bfg \right]^T_0\ ,
\eqn\idt
$$
where we have used the inner product 
$\bfa\cdot \bfb = a^i b^i$.
If we choose for the functions the solutions
$\bff = \bfu_k$, $\bfg = \bfv_l$ 
of the Jacobi equation (\jacobieq), 
then from the boundary 
conditions, (\vsol), (\usol), (\ucaust) and (\unorm),
we find that in matrix form the relations (\idt) 
become
$$
V^T(T) P(T) {\dot U}(T) - {\dot V}^T(T) P(T) U(T) 
+ V^T(T)\bigl\{Q^T(T) - Q(T)\bigr\} U(T) =  P(0) {\dot U}(0)\ .
\eqn\ida
$$
We may also choose $\bff = \bfu_k$ and $\bfg = \bfu_l$ to get
$$
U^T(T)  P(T) {\dot U}(T) - {\dot U}^T(T) P(T) U(T)
+ U^T(T)\bigl\{Q^T(T) - Q(T)\bigr\} U(T) = 0\ .
\eqn\idb
$$
In particular, for $l > f$ and $k \le f$ the $(lk)$-component
of the identities (\idb) reads
$$
{\bfu}_l(T) \cdot P(T)\, \dot{\bfu}_k(T) = 0\ .
\eqn\idc
$$
In the non-caustics case 
where one has $U(T) = 1$,
one can eliminate
$Q(T)$ from the identity (\ida) by using (\idb)
to find
$$
\left\{ - {\dot V}^T(T) + V^T(T) {\dot U}^T(T)
\right\} P(T) = P(0) {\dot U}(0)\ .
\eqn\idd
$$

Let us recall the general definition of the 
Jacobi fields.  Denote by 
$\bar{\bfx}(\bfp, t)$ the classical
solution with 
$$
\bar{\bfx}(\bfp, 0) = \bfa, \qquad 
{{\partial L[\bar{\bfx}]}\over
{\partial \dot{\bar{\bfx}}}}\bigg\vert_{t = 0} 
= \bfp\ ,
\eqn\jacdef
$$
{}for some given $\bfa$, $\bfp \in \R^d$.
Then the Jacobi fields are given by
$$
J_{ik}(t) := {{\partial \bar{x}^i(p, t)}\over{\partial p_k}}\ .
\eqn\jacobi
$$
{}For our quadratic system (\lagrangian) we have 
the classical 
solution in the form (\general)
and $p_k$ is given by 
$P_{ki}(0)\, \dot{\bar{x}}^i(0) 
+ a^i\, Q_{ik}(0)$.  From these 
the Jacobi fields are found to be
$$
J(t) = U(t)\dot{U}^{-1}(0) P^{-1}(0)\ .
\eqn\jacb
$$
This shows manifestly that $J(t)$ fulfills 
the Jacobi equation (\jacobieq) with the initial condition
$J(0) = 0$.

At this point we remark that, as can be seen from 
the explicit form of 
the classical solution (\general) 
the classical action for the solution is at most
quadratic in the boundary values,
$$
I[\bar{\bfx}] = W_{ij}a^i a^j + X_{ij} a^i b^j + Y_{ij}b^ib^j
+ E_ia^i + F_ib^i + G,
\eqn\clac
$$
where $W_{ij}, X_{ij}, \dots, G$ are some functions of $T$.   
(Note that in (\clac)
the summation for $a^i$ over $i$ runs from 
$1$ to $d$ whereas for $b^i$ it runs 
only from $f+1$ to $d$ due to
the choice in (\asol).) For instance, 
in terms of our Jacobi fields, $X_{ij}$, $i = 1, \ldots, d$,
$j = f+1, \ldots, d$, are given by
$$
\eqalign{
X_{ij} 
&= {1\over 2}\biggl[ - P_{ik}(0) {\dot u}^k_j(0)
+ \Bigl\{{\dot v}^k_i(T) 
- \sum_{l = f+1}^d v^l_i(T) {\dot u}^k_l(T) \Bigr\}
P_{kn}(T) u^n_j(T)
\biggr] \cr
& + {1\over 2}\Bigl\{v^k_i(T) 
- \sum_{l = f+1}^d v^l_i(T) u^k_l(T) \Bigr\}
\biggl[ P_{kn}(T) {\dot u}^n_j(T) 
+ \Bigl\{Q_{kn}(T) + Q_{nk}(T)\Bigr\} u^n_j(T)
\biggr].
}
\eqn\vv
$$ 
In particular, for non-caustics case 
the identity (\idd) can be used to simplify (\vv)
into
$$
X = - {1 \over 2} \left\{ P(0) {\dot U}(0)
+ \left[- {\dot V}^T(T) + V^T(T) {\dot U}^T(T)\right] 
P(T) \right\}
= - P(0) {\dot U}(0)\ .
\eqn\vvb
$$
Using (\vvb) and (\jacb), one can immediately confirm 
the relation,
$$
- J(T)\, X = - X\, J(T) = 1,
\eqn\invse
$$
which follows from the general definition (\jacdef).

\bigskip
\noindent
{\bf 2.2. Quantum caustics}

We now move on to quantum mechanics and consider
the problem corresponding to the classical 
Dirichlet problem (\bc), that is, we wish to find 
the transition kernel $K(\bfb, T; \bfa, 0)$ whose
path-integral expression is
$$
K (\bfb,T;\bfa,0)
  = \int_{\exbfx(0)=\exbfa}^{\exbfx(T)=\exbfb} 
{\cal D} \bfx \, e^{{i\over\hbar} I[\exbfx]}\ .
\eqn\pathint
$$
To this end, as we did for
the one-dimensional case,
we first decompose any path $\bfx(t)$ with
the boundary values (\bc) as
$$
        \bfx(t) = \bar{\bfx}(t) + \bfrho(t) + \bfeta(t)\ .
\eqn\dec
$$
Here
$\bar{\bfx}(t)$ is a classical path starting from
$\bar{\bfx}(0) = \bfa$ 
and ending at a point in the caustic surface
(\cstsurface), and for definiteness 
we choose to be the one (\oursol)
mentioned earlier.  The function $\bfrho(t)$  
is a compensating function satisfying
$ \bfrho(0) = \bfzero$ and 
$$
\bfrho(T) = \bfb - \bar{\bfx}(T) 
= U^{\bot}(T) (\bfb - \bfh(\bfa)) \ ,
\eqn\compensate
$$
which is designed to fill the gap 
between the given endpoint $\bfb$ for the transition and
the actual endpoint $\bar{\bfx}(T)$ of the classical 
solution.  Note that $U(T) \bfrho(T) = 0$ implies 
$\rho^i(T) = 0$ for $i = f+1, \ldots, d$.

The final piece $\bfeta(t)$ in (\dec) representing 
fluctuations fulfills 
$ \bfeta(0) = \bfeta(T) = 0$ and may be expanded
as $ \bfeta(t) = \sum_n a_n \bfchi_n(t)$
in terms of the orthonormal vector-valued eigenfunctions
$\{ \bfchi_n \}$ associated with the self-adjoint operator
$\Lambda$:
$$
\Lambda(t)\, \bfchi_n(t) = \lambda_n \, \bfchi_n(t)\ ,
\eqn\eigeneq
$$
with
$$
  \bfchi_n(0) = \bfchi_n(T) = 0\, ; \qquad
  \int_0^T dt\, \bfchi_n(t)\cdot \bfchi_m(t) = \delta_{nm} .
\eqn\normalization
$$
The action for 
an arbitrary path $\bfx(t)$ then becomes
$$
\eqalign{
I[\bfx] &=  I[ \bar{\bfx} + \bfrho + \bfeta ] \cr
     &= I[ \bar{\bfx} + \bfrho]
  +
        \frac{1}{2} \sum_n \lambda_n a_n^2
+ \sum_n \lambda_n a_n \int_0^T dt \, 
  \bfrho(t)\cdot \bfchi_n(t) \cr
& \qquad + \bfrho(T)\cdot P(T)
  \sum_n a_n \dot{\bfchi}_n(T).
}
\eqn\claction
$$

Notice that, 
when caustics occur, the solutions $\{ \bfu_k \}$ for 
$k = 1, 2, \ldots, f$ may be chosen 
such that they are orthonormal each other, 
$\int_0^T dt\, \bfu_k(t)\cdot \bfu_l(t) = \delta_{kl}$,
by performing linear
transformations among themselves.
This implies that $\{ \bfu_k \}$ 
is just the set of $f$ zero modes
$\{ \varphi_k \}$ with $\lambda_k = 0$ in the orthonormal
set of eigenfunctions.  Keeping this in mind, 
by change of the measure ${\cal D}\bfx = {\cal D}\bfeta   
\propto \prod_n da_n$,
we carry out the path-integration (\pathint) to obtain
$$
K (\bfb,T;\bfa,0)
   =  \left(\frac{2 \pi}{i}\right)^{f/2} {\cal N}
        \left[ \mathop{{\prod}'}_n \lambda_n\right]^{ -1/2 }
       \prod_{k = 1}^f \delta( \bfrho(T)\cdot P(T)
        \, \dot{\bfu}_k(T) )\, 
        e^{{i\over\hbar}I[\bar{\exbfx} + \exbfrho]}\ ,
\eqn\no
$$
where ${\cal N}$ is a normalization constant and 
the prime in $\mathop{{\prod}'}_n$ indicates
that the zero modes are omitted in the product.
If we denote by 
$\det' M$ the minor corresponding to the first $f \times f$
part of a $d \times d$ matrix $M$, then from the property
(which we shall show later),
$$
{\det}'(P(T)\dot{U}(T)) \ne 0,
\eqn\detcon
$$ 
we find
$$
\prod_{k = 1}^f 
\delta( \bfrho(T)\cdot P(T)
        \, \dot{\bfu}_k(T) )
= \vert{\det}'(P(T){\dot U}(T))\vert^{-1}
\prod_{i = 1}^f \delta( \rho^i(T) )\ .
\eqn\deleq
$$
Thus, if we let $m(T)$ 
be the Morse index of the operator
$\Lambda$ (associated with the period $[0, T]$) which gives
the number of non-positive modes $\lambda_n \le 0$ 
in (\eigeneq),
and combine (\deleq) with (\compensate), we get the kernel
in the polar form, 
$$
K (\bfb,T;\bfa,0) = R(T)\, \prod_{i = 1}^f 
\delta\left(
\left[ U^{\bot}(T) (\bfb - \bfh(\bfa)) \right]^i 
\right)\,
e^{ i \Theta(\exbfb,T;\exbfa,0) }\ .
\eqn\mkrn
$$
The phase part 
is given by 
$$
\Theta(\bfb,T;\bfa,0) := {1\over \hbar}
I[\bar{\bfx}] - {{\pi}\over 2} m(T) + \gamma \ ,
\eqn\phase
$$
where $\gamma$ is a constant independent of $T$, and 
we have replaced $I[\bar{\bfx} + \bfrho]$ 
with $I[\bar{\bfx}]$ 
under the presence of the delta-functions in (\mkrn).  
The result (\mkrn) shows that, when caustics occur, allowed
transitions are those satisfying $\bfrho(T) = \bfzero$, that is, 
those whose boundaries admit classical
solutions.  In other words, classically forbidden
processes remain to be forbidden even quantum mechanically.

The key ingredient for 
determining the modulus part $R(T)$ of the kernel (\mkrn)
is the unitarity relation,
$$
  \prod_{i = 1}^d \delta (a^i - c^i)
= \int \prod_{i = 1}^d  db^i\,
K^* (\bfb,T;\bfc,0) \, K (\bfb,T;\bfa,0)\ .
\eqn\unirel
$$
Plugging (\mkrn) into (\unirel) and noticing 
$$
\Theta(\bfb,T;\bfa,0) -  \Theta(\bfb,T;\bfc,0) 
= {1 \over \hbar} \sum_{j = f+1}^d X_{ij}(a^i - c^i) b^j
+ \Phi(\bfa-\bfc)\ ,
\eqn\no
$$ 
where $\Phi(\bfa-\bfc)$ stands for the
terms which are independent of $\bfb$ and vanish at 
$\bfa = \bfc$,
we observe that the r.h.s.~of (\unirel) becomes
$$
\eqalign{
&R^2(T) \int \prod_{j = 1}^d db^j\,
\prod_{l = 1}^f 
\delta\left(
\left[ U^{\bot}(T) (\bfb - \bfh(\bfa)) \right]^l
\right)
\cr
&\qquad\qquad \times 
\prod_{m = 1}^f 
\delta\left(
\left[ U^{\bot}(T) (\bfb - \bfh(\bfa)) \right]^m 
\right)
\, e^{ i\Theta(\exbfb,T;\exbfa,0) 
- i\Theta(\exbfb,T;\exbfc,0)  } 
\cr
&= R^2(T) \int \prod_{j = f+1}^d db^j\,
\prod_{m = 1}^f 
\delta\left(
\left[ U^{\bot}(T) (\bfh(\bfa) - \bfh(\bfc)) \right]^m 
\right)
e^{{i\over\hbar} \sum_{j = f+1}^d X_{ij}(a^i - c^i) b^j 
 + i\Phi(\exbfa-\exbfc)} 
\cr
&= R^2(T)
\prod_{m = 1}^f
\delta\left(
\left[ U^{\bot}(T) (\bfh(\bfa) - \bfh(\bfc)) \right]^m
\right)
(2\pi\hbar)^{d - f}
\prod_{j = f+1}^d
\delta\left(X_{ij} (a^i - c^i)\right)\, 
e^{i\Phi(\exbfa-\exbfc)}\ .
}
\eqn\no
$$
We then use $\bfh(\bfa) - \bfh(\bfc) 
= V(T)(\bfa - \bfc)$ together with 
the matrix $Z$ defined by 
$$
Z_{ij} = 
\cases{ \left\{\delta_n^i - u^i_n(T)\right\} v^n_j(T), 
           &for $\quad j = 1, 2, \ldots, f$; \cr
  X_{ji} , &for $\quad j = f+1, f+2, \ldots, d$, \cr}
\eqn\zedd
$$
to rewrite the unitarity relation (\unirel) as
$$
\eqalign{
\prod_{i = 1}^d \delta (a^i - c^i)
&= (2\pi\hbar)^{d - f} R^2(T)\prod_{i = 1}^d 
\delta\left(Z_{ij}(a^j - c^j)\right)\,
e^{i\Phi(\exbfa-\exbfc)} \cr
&= (2\pi\hbar)^{d - f} R^2(T) \, \vert\det Z\vert^{-1}\, 
\prod_{i = 1}^d \delta (a^i - c^i)\ .
}
\eqn\unib
$$
{}From this the modulus part is found to be
$$
R(T) 
= (2\pi\hbar)^{-{{d - f}\over{2}}} \sqrt{\vert \det Z\vert}\ .
\eqn\modul
$$
We here point out that the fact
$\det Z \ne 0$ can be shown directly, but it is also obvious from 
the observation that otherwise
the r.h.s.~of (\unib) does not match
the l.h.s.~due to the
difference in the structure of delta-functions.

It remains to determine the constant $\gamma$ in the
phase factor (\phase).  We do this by looking at
the initial condition for the kernel,
$$
\lim_{T \to 0+} K (\bfb,T;\bfa,0) 
= \prod_{i = 1}^d \delta(b^i - a^i)\ .
\eqn\incon
$$  
Note that
the limit $T \rightarrow 0$ cannot
be taken for our kernel (\mkrn), since caustics take place only
at finite $T$.  However, the formal limit still makes sense
if we put $U(T) \rightarrow U(0) = 0$ 
and $V(T) \rightarrow V(0) = 1$ in 
(\mkrn) and thereby isolate the delta-functions
$
\prod_{i = 1}^f \delta 
([ U^{\bot}(T) (\bfb - \bfh(\bfa))]^i) 
\rightarrow
\prod_{i = 1}^f \delta (b^i - a^i)
$
so that the rest of the kernel describes the
transition on the caustics plane.   
To use (\incon), we also have to evaluate
the classical action in the phase factor for 
the solution (\general),
but for our purpose we only need the asymptotic form 
of the solution,
$$
\bar{\bfx}(t) = \bfa 
+ (\bfb - \bfa){t \over T} + {\cal O}(T)\ ,
\eqn\asymp
$$
which leads to the classical action, 
$$
\eqalign{
I[\bar{\bfx}] &= 
{1\over {2T}} (\bfb - \bfa)\cdot P(0) (\bfb - \bfa) \cr
& \quad + {1\over 4} (\bfb - \bfa)\cdot 
\dot P(0) (\bfb - \bfa) 
+ {1\over 2} (\bfb + \bfa)\cdot Q(0) (\bfb - \bfa)
+ {\cal O}(T)\ .
}
\eqn\caaymp
$$
{}From (\zedd) we observe that $\det Z$ reduces
in the limit to $\det''X$ which is given by 
the minor corresponding to the last $(d-f) \times (d-f)$
part of the matrix $X$.  Thus from the classical action
(\caaymp) we find
$$
\vert \det Z \vert 
= T^{-(d-f)} {\det}''P(0) + {\cal O}(T^{-(d-f)+1})\ .
\eqn\no
$$
With the help of the identity
$
\lim_{\epsilon \to 0}(2\pi i\epsilon)^{-n/2}\, 
e^{i{\exbfx}\cdot A {\exbfx} /2\epsilon} 
= (\det A)^{-1/2} \delta^{(n)}({\bfx})
$ 
valid for 
an $n$-dimensional vector $\bfx$ and an $n \times n$ matrix $A$, 
together with the property $m(T) = 0$ for $T \rightarrow 0$, 
we can readily evaluate the (formal) limit of the kernel (\mkrn)
to get
$$
\lim_{T \to 0+} K (\bfb,T;\bfa,0) 
= i^{\frac{d-f}{2}}\, e^{i\gamma}
\prod_{i = 1}^d \delta(b^i - a^i)\ .
\eqn\no
$$  
Comparing with (\incon) we find that the constant $\gamma$ 
is determined by 
$e^{i\gamma} = i^{-(d-f)/2}$.  Having found both
the modulus and the phase part, we obtain 
the closed form of the transition kernel 
on caustics:
$$
K (\bfb,T;\bfa,0) = 
(2\pi i\hbar)^{-{{d - f}\over{2}}} 
\sqrt{\vert \det Z\vert}\, \prod_{i = 1}^f 
\delta\left(
\left[ U^{\bot}(T) (\bfb - \bfh(\bfa)) \right]^i 
\right)\,
e^{ {i\over\hbar} I[\bar{\exbfx}] - {{i\pi}\over 2} m(T)}\ .
\eqn\finalkrl
$$

In particular, for the full caustics case $f = d$
the kernel reduces to
$$
K (\bfb,T;\bfa,0) = \sqrt{\vert \det V(T)\vert}\, 
\prod_{i = 1}^d 
\delta\left( b^i - h^i(\bfa) \right)\,
e^{ {i\over\hbar} I[\bar{\exbfx}] 
- {{i\pi}\over 2} m(T) }\ .
\eqn\fullc
$$
On the other hand, in the non-caustics case $f = 0$ 
the matrix $Z$ becomes the Van Vleck matrix 
$X$ with $X_{ij} 
= \partial^2 I[\bar{\bfx}]/\partial a^i \partial b^j$, 
and hence the kernel,
$$
K (\bfb,T;\bfa,0) =  (2\pi i\hbar)^{-{{d}\over{2}}} 
               \sqrt{\vert \det X \vert}\, 
e^{ {i\over\hbar} I[\bar{\exbfx}] 
- {{i\pi}\over 2} m(T) }\ ,
\eqn\nonc
$$
is indeed the standard one (\vvformula) for 
the quadratic system (with the phase correction by
the Morse index included) --- 
here we derived it directly from 
the unitarity relation (\unirel) and the initial 
condition (\incon) without relying on
involved procedures, such as one using 
discretized multiple 
integrations and recursion relations
employed in the literature.
We also mention that, with the generic 
Jacobi fields $J(t)$ 
satisfying (\invse), 
the transition kernel (\nonc) becomes the generalized
Gel'fand-Yaglom formula obtained previously [\LS].

Before closing this section, we wish to 
prove the property (\detcon) used to obtain the kernel (\mkrn).
To this end, first we assume that
(\detcon) does not hold.  Then we can find a
linear combination 
$\bfu(t) := \sum_{k = 1}^f c_k \bfu_k(t)$ out 
of the solutions $\{ \bfu_k; k = 1, \ldots, f \}$ such that 
$P(T)\dot{\bfu}(T) =  (0, \ast)^T$, 
where `0' is an $f$-dimensional 
null vector.  But since (\idc) 
is equivalent to $U^T(T)\, P(T)\dot{\bfu}_k(T) = 0$ for 
$k = 1, \ldots, f$, we deduce
$U^T(T)\, P(T)\dot{\bfu}(T) = 0$.  This 
suggests that the last $d-f$ components of 
the vector $P(T)\dot{\bfu}(T)$ also vanish identically,
{\it i.e.}, $P(T)\dot{\bfu}(T)$ is actually a null vector.
Then from the positive definiteness of $P(T)$
we find $\dot{\bfu}(T) = 0$.  Combined with
$\bfu(T)=0$ which follows from the caustics conditions, 
we come to the conclusion that $(\bfu(T), \dot{\bfu}(T)) = 0$
as a $2d$-dimensional vector.  This however contradicts with
the non-triviality of the solutions 
$\{ \bfu_k\}$, $\{ \bfv_k\}$ at $t = T$,
$$
\det \pmatrix{
U(T) & V(T) \cr
\dot{U}(T) & \dot{V}(T) \cr} 
\ne 0\ ,
\eqn\nnt
$$
which ensures that, 
as a set of $2d$-dimensional vectors,
$(\bfu_k(T), \dot{\bfu}_k(T))$ 
(together with $(\bfv_k(T), \dot{\bfv}_k(T))$) for 
$k = 1, \ldots, d$ form a complete basis.
We therefore see that our assumption is wrong,
proving (\detcon) as claimed.

\ve

\secno=3 
\meqno=1


\centerline{\bf 3. Caustics 
under External Electromagnetic Fields}
\medskip

In this section we provide two examples to illustrate
caustic phenomena for a particle in three dimensions.
The first example discusses partial caustics $f=2$,
which however can be more naturally viewed 
as two-dimensional full 
caustics plus a decoupled motion in the third dimension.
The second example shows non-trivial partial caustics $f=1$.
Electromagnetic fields are used as driving forces acting on
the test particle, and the two examples show that caustics 
with different focal dimensions are 
possible for different field 
configurations.

\bigskip
\noindent
{\bf 3.1. Full caustics}

Let us consider a point particle with mass $m$ and charge
$q$ subject to a non-static but uniform electric field
${\bfE_{\rm ext}} = {(E_x(t),E_y(t),E_z(t))}^T$ 
and, assuming an
appropriate background current, a static uniform 
magnetic field 
${\bfB_{\rm ext}} = {(0,0,B)}^T$. These 
fields can be derived from 
a vector potential $\bfA$ in the symmetric gauge,
$ {\bfA} ={B \over 2}{(-y, x, 0)}^T$, and a scalar
potential 
$\phi$ is given by $\phi = {\bfx}\cdot{\bfE_{\rm ext}}$.
Noting the position of the particle by
${\bfx} = {(x, y, z)}^T$ its Lagrangian reads
$$
  L_0 := {m \over 2}{{\dot{{\bfx}}}^2}
        +q{\dot{{\bfx}}}\cdot{{\bfA}}
        -q\phi \,.
\eqn\L
$$
Although essentially the same model has been considered in
[\FC], we re-examine it in detail to
illuminate the phenomenon of caustics from our point of view.

Due to the special field configuration the Lagrangian (\L)
splits into two parts,
$$
  L_0 = L + L_z \,,
\eqn\Ldec
$$
of which $L_z = {m \over 2}{{\dot z}^2}-q {E_z}(t)z$
describes an accelerated motion in the $z$-direction 
whereas 
$$
  L = {m \over 2}\left( {\dot{\bfr}}^2
                       +\omega{\bfr}\cdot\Omega\dot{\bfr}
                 \right)
     -q{\bfE}\cdot{\bfr} \,, 
\eqn\LagL
$$
governs the motions in the $xy$-plane where we use
${\bfr} = (x, y)^T$ and ${\bfE}=({E_x},{E_y})^T$.
We have also introduced the cyclotron angular velocity
$\omega := qB/m$ and the skew matrix
$\Omega = \left(\matrix{ 0 & 1 \cr -1 & 0 \cr}\right)$.
By splitting off the $z$-component in (\Ldec) we obtain the 
system (\LagL) on 
which we shall concentrate in the following.
 
As is evident from the discussion in the previous Section, the
intrinsic behavior of the system on caustics 
does not depend on
terms linear in the coordinates, $-q{\bfE}\cdot{\bfx}$.
Without this term the system exhibits the discrete 
symmetry under the `parity' conjugation
$(x(t),\omega) \leftrightarrow (y(t),-\omega)$,
which has important consequences on the caustics. 

The equation of motion for a classical path $\bar{\bfr}$ reads
$$
  \Lambda\,{\bar{\bfr}} + {\bfS} = 0 \,,
\eqn\EQM
$$
in which the differential operator $\Lambda$ 
and the nonlinear term ${\bfS}$ are given by
$$
   \Lambda = -m {{d^2}\over{d {t^2}}} 1
            + m\omega\Omega{d \over{dt}} \,,
\qquad 
   {\bfS}(t) = -q {\bfE}(t)\ ,
\eqn\DEFO
$$
where $1$ denotes a two by two unit matrix.
To provide a special solution of 
(\EQM) we focus on the first time
derivative ${\bfy}(t) := \dot{\bar{\bfr}}(t)$, which fulfills
$$
  m(\ddot{\bfy} + \omega^2 \bfy)
  + q(\omega\Omega{\bfE}+\dot{\bfE}) = 0 \,.
\eqn\eqmi
$$
This equation is solved by
$$
  {\bfy}(t) = - {q\over{m\omega}}\int^t_0 dt'\sin\omega(t-t')\,
                \{\omega\Omega{\bfE}(t')+\dot{\bfE}(t')\}
               +{\bfg}(t) \,,
\eqn\ygensol
$$
wherein ${\bfg}(t)$ stands for solutions to the
homogeneous part of (\eqmi) given by a linear combination
of $\sin \omega t$ and $\cos \omega t$. 
By integrating the solution (\ygensol) and thereby
imposing the boundary condition 
$\bar{\bfs}(0) = {\bfzero}$ we obtain the special solution,
$$
  {\bfs}(t) = 
  {q\over{m\omega}}\Omega
  \int^t_0 dt'\left(e^{\Omega\omega(t-t')}-1 \right)E(t')\ .
%
 \eqn\specS
$$
On the other hand, 
the solution space of the homogeneous part of (\EQM)
is formed by the four independent functions,
$$
  \left(\matrix{1 \cr 0}\right) \,,\quad 
  \left(\matrix{0 \cr 1}\right) \,,\quad
  \left(\matrix{\sin\omega t \cr  \cos\omega t}\right) \ ,
  \quad
  \left(\matrix{\cos\omega t \cr -\sin\omega t}\right) \ .
\eqn\fourhom
$$
In particular, the Jacobi fields $\bfv_1$ and
$\bfv_2$ which satisfy (\vsol) are given by the
first two constant vectors, 
respectively, while 
$\bfu_1$ and $\bfu_2$ are given by linear combinations of
the four.  
The general solution for the equation of motion (\EQM) is 
then obtained by (\general) (with $\bar{\bfx}$ replaced by
$\bar{\bfr}$).

In order to establish the conditions for caustics,
let us consider the eigenvalue equation 
(\eigeneq) with (\normalization) for the present case.
Although the eigenfunctions of 
this equation can be found in the
usual manner, for completeness let us briefly comment
on their derivation.  
First, for the eigenfunctions we make the ansatz
${\bfchi}_n(t) = {\bfc}\, e^{i\xi t}$,
where $\bfc$ is a constant complex two-vector and $\xi$ a
complex number.  From (\eigeneq) one 
deduces the coefficient equation
$[(m{\xi^2} - {\lambda_n})1 + im\xi\Omega] {\bfc} = 0$, 
which in turn implies that $\xi$ can take one of 
the following four values,
$$
\matrix
{ \;\,{ \xi_1 } = {-\omega + 
\sqrt{\omega^2+(4\lambda_n/m)} \over 2},
& \;\;{ \xi_2 } = {-\omega - 
\sqrt{\omega^2+(4\lambda_n/m)} \over 2},\cr
     { \xi_3 } = { \omega + 
\sqrt{\omega^2+(4\lambda_n/m)} \over 2},
& { \xi_4 } = { \omega - 
\sqrt{\omega^2+(4\lambda_n/m)} \over 2},\cr
}
\eqn\no
$$
to each of which there corresponds a fixed vector
${\bfc} = {\bfc}_i$ given by
${\bfc}_1 = {\bfc}_2 = (i,\; 1{)^T}$ and
${\bfc}_3 = {\bfc}_4 = (-i,\; 1{)^T}$, respectively.
The general solution to (\eigeneq) reads ($a_i$ complex)
$$
  \left(\matrix{ i\cr1\cr}\right)
  ( a_1 e^{i\xi_1 t} + a_2 e^{i\xi_2 t} )
  +
  \left(\matrix{-i\cr1\cr}\right)
  ( a_3 e^{i\xi_3 t} + a_4 e^{i\xi_4 t} ) \,,
\eqn\no
$$
upon which the boundary condition in (\eigeneq) 
is to be imposed.
This leads to $a_1 + a_2 = 0$, $a_3 + a_4 = 0$, 
$e^{i\xi_1 T} = e^{i\xi_2 T}$, and
$e^{i\xi_3 T} = e^{i\xi_4 T}$, from which we infer
$m\omega^2+4\lambda_n > 0$ and deduce the eigenvalues
$$
  \lambda_n = -{{m\omega^2}\over 4} 
+ m\left({{n\pi}\over T}\right)^2
\eqn\Eigenvalue
$$
{}for positive integer $n$. There exist two real 
eigenfunctions to each
eigenvalue, namely,
$$
  {\bfchi}_n{}^{\!(+)}(t) =
    \sqrt{T\over2} \sin\left({{n\pi}\over T} t \right)
    \left(\matrix{\sin\left({{\omega t}\over 2}\right)\cr
                  \cos\left({{\omega t}\over 2}\right)}\right)
\eqn\uplus
$$
and
$$
  {\bfchi}_n{}^{\!(-)}(t) =
    \sqrt{T\over 2} \sin\left({{n\pi}\over T} t \right)
    \left(\matrix{-\cos\left({{\omega t}\over 2}\right)\cr
       \;\;\;\sin\left({{\omega t}\over 2}\right)}\right)\,.
\eqn\uminus
$$
One can check that these two series of eigenfunctions
(\uplus) and (\uminus) are indeed orthonormal, 
$\int^T_0 dt {\bfchi}_n^{(l)}\cdot {\bfchi}_m^{(l')} =
\delta_{nm}\delta^{l l'}$ and complete.
The twofold degeneracy comes from the discrete 
symmetry of the Lagrangian (\LagL).  
Clearly, zero modes $\lambda_n = 0$ appear if 
$$
\omega = {\omega_n} \qquad \hbox{where} \quad  
\omega_n := \frac{2n\pi}{T} ,
\eqn\caustcondition
$$
in which case caustics occur.  Note that at any of the above
$\omega_n$ for some positive integer 
$n$, there appear two zero modes 
due to the degeneracy.  We thus find that, as far as the
the motions on the $xy$-plane are concerned, we have
the full caustics, $f = 2$.   

Let us turn our attention 
to quantum mechanics and obtain the transition
amplitude on the caustics at $\omega = \omega_n$.
Being the full caustics, the amplitude can be obtained 
immediately by using the general 
formula for the kernel (\fullc),
where now we have $\vert \det V(T) \vert = 1$ and 
$\bfh(\bfa) = \bfa + \bfs(T)$ with $\bfs(T)$
being the endpoint value of the special solution (\specS).
We also notice from (\Eigenvalue) that
the Morse index 
is given by $m(T) = 2n$ 
(where the factor 2 comes from 
the twofold degeneracy) and the classical action reads
$$
  I[\bar{\bfr}]
  = -{q\over 2}{\bfa} \cdot
     \int^T_0 dt \left(e^{\Omega\omega(T-t)}+1\right){\bfE}(t) 
  +{\bfs}(T) \cdot {m\over 2}\dot{\bfs}(T)
     -{q\over 2} \int^T_0 dt\, {\bfs}(t) \cdot {\bfE}(t) \ .
\eqn\no
$$
Combining these, for $\omega = \omega_n$ we obtain 
$$
    K({\bfb},T;{\bfa},0)
       = (-1)^n\, \delta^{(2)}(\bfb - \bfa - \bfs(T))\,
         e^{{i\over\hbar}I[\bar{\exbfr}]} \ .
\eqn\knl
$$

So far we have discussed the two 
dimensional subsystem governed by the
Lagrangian $L$, which reveals full 
caustics at $\omega = \omega_n$.
When the total 
system $L_0$ in (\Ldec) is considered,
these caustics are only partial, since the motion in the
$z$-direction is free of caustics.  
This fact is also expressed in
the total transition amplitude from the initial position
$({a_x},{a_y},{a_z})^T$ to the final 
one $({b_x},{b_y},{b_z})^T$, 
which is the
product of the kernel 
(\knl) and the contribution coming from the
$z$-component
$$
  K({\bfb},T;{\bfa},0) \times {K_z}({b_z},T;{a_z},0),
\eqn\totknl
$$
where ${K_z}({b_z},T;{a_z},0)$ can be calculated as
$$
    {K_z}({b_z},T;{a_z},0)
    = \sqrt{m \over 2\pi i\hbar T} 
      \, e^{{i\over\hbar}I_z}\ ,
\eqn\zedknl
$$
with
$$
\eqalign{
    {I_z}
    & = {m \over 2T}
        \left[({b_z}-{a_z})^2 
        + {2 \over m} {\int^T_0}dt {E_z}(t)
              \{{a_z} T + t({b_z}-{a_z})\}\right.\cr
    &\qquad\qquad
        \left. -{2 \over m^2}{\int^T_0}dt {\int^t_0}dt'
                {E_z}(t){E_z}(t')(T- t)t'
        \right] \,.\cr
  }
\eqn\zedaction
$$
Since the $z$-component $b_z$ of the endpoint can take an
arbitrary value, depending on the initial momentum in this
direction, the set of image points of an initial point ${\bfa}$
makes up a straight line in the $z$-direction.
The transition kernel obtained here, 
(\totknl) with (\knl), (\zedknl) and (\zedaction),
agrees with the one [\FC] obtained 
previously by a different method.

\bigskip
\noindent
{\bf 3.2. Partial caustics}

The above example, when viewed as that of partial caustics,
is rather trivial due to the decoupling of the $z$-component. 
However, partial caustics do not always have 
such a trivial situation. 
To examine a non-trivial case of partial caustics let us
consider the following Lagrangian in three dimensions,
$$
  L = {m \over 2} (\dot{x}^2 + \dot{y}^2 + \dot{z}^2)
     +{{qB} \over 2} (x \dot{y} - y \dot{x})
     -\alpha q y z \,.
\eqn\Lqa
$$
It describes a point particle of mass $m$ and charge
$q$ moving in a static uniform 
magnetic field in the $z$-direction
${\bfB_{\rm ext}} 
= (0,0,B{)^T}$ and a static but non-uniform electric field
${\bfE_{\rm ext}} = -\alpha(0,z,y{)^T}$, $\alpha \neq 0$.
Introducing $\omega = qB/m$ and 
$\gamma = q\alpha/m$ the equation of
motion reads
$$
   \ddot{{\bfx}}(t)
  +\left(\matrix{ 0&-\omega&0 \cr 
    \omega&0&0 \cr 0&0&0 \cr}\right)
   \dot{{\bfx}}(t)
  +\left(\matrix{ 0&0&0 \cr 0&0&\gamma \cr 
     0&\gamma&0 \cr}\right)
   {\bfx}(t) = {\bfzero} \,.
\eqn\EQMT
$$
We note that, like the previous example, 
the Lagrangian and the equation of motion are
invariant under the `parity' transformation,
$$
  (x,y,z,\omega,\gamma) \rightarrow
  (x,-y,z,-\omega,-\gamma) \,.
\eqn\no
$$

In order to construct classical solutions we make the
ansatz ${\bfx}(t) = {\bfh}e^{i\lambda t}$ (${\bfh}$ a
complex constant vector and $\lambda$ a complex number) 
and substitute it into (\EQMT), yielding
$$
  \left[
    {\lambda^2} 
   \left(\matrix{ 1& 0& 0 \cr 0&1&0 \cr 0&0&1 \cr}\right)
   -i\lambda
    \left(\matrix{ 0&-\omega&0 \cr \omega&0&0 
       \cr 0&0&0 \cr}\right)
   -\left(\matrix{ 0&0&0 \cr 0&0&\gamma \cr 
    0&\gamma&0 \cr}\right)
  \right]
  {\bfh} = {\bfzero} \ .
\eqn\EQH
$$
These equations possess the pair of solutions
$(\lambda,{\bfh})$ given by
$$
  \eqalign{
    \lambda_1 & = 0 \,;\;\; \qquad
    {\bfh}_1 = \left(\matrix{1\cr0\cr0\cr}\right) , 
     \qquad\quad\;\!
    {\bfh}'_1= \left(\matrix{\gamma t\cr0\cr 
     -\omega\cr}\right) ,
    \qquad \hbox{(double root)}
  \cr
    \lambda_2 & =  \xi \,;\;\; \qquad
    {\bfh}_2 = \left(\matrix{-i\omega\cr\xi\cr
     \gamma/\xi\cr}\right) ,
    \qquad\lambda_3 = -\xi \,; \qquad
    {\bfh}_3 = \left(\matrix{+i\omega\cr\xi\cr
      \gamma/\xi\cr}\right)
              = {\bfh}_2^* \,,%
  \cr
    \lambda_4 & =  i\eta \,;\; \qquad
    {\bfh}_4 = \left(\matrix{+\omega\cr-\eta\cr
     \gamma/\eta\cr}\right) ,
    \qquad\lambda_5 = -i\eta \,; \qquad
    {\bfh}_5 = \left(\matrix{-\omega\cr-\eta\cr
      \gamma/\eta\cr}\right) ,
  \cr}
\eqn\no
$$
where we have used
$$
  \xi   = \sqrt{\beta+1 \over 2} \; \omega    \,, \quad
  \eta  = \sqrt{\beta-1 \over 2} \; \omega     \,, \quad
  \beta = \sqrt{1+4{\gamma^2\over w^4}} > 1 \ .
\eqn\no
$$
By taking linear combinations of the 
complex-valued solutions
${{\bfh}_j}e^{i{\lambda_j} t}$ for $j=2$ 
and 3 and likewise for
$j=4$ and 5 we obtain the following 
real solutions besides
${\bfh}_1$ and ${\bfh}'_1$,
$$
\eqalign{
    {\bff}_2(t)  =
    \left(\matrix{ \omega     \sin\xi t \cr
                   \xi        \cos\xi t \cr
                   (\gamma/\xi) \cos\xi t \cr }\right) ,
  & \qquad
    {\bff}_3(t)  =
    \left(\matrix{ \omega     \cos\xi t \cr
                  -\xi        \sin\xi t \cr
                  -(\gamma/\xi) \sin\xi t \cr }\right) ,
  \cr
    \qquad\qquad\qquad{\bff}_4(t)  =
    \left(\matrix{-\omega     \sinh\eta t \cr
                  -\eta       \cosh\eta t \cr
                   (\gamma/\eta)\cosh\eta t \cr }\right) ,
  & \qquad
    {\bff}_5(t)  =
    \left(\matrix{-\omega     \cosh\eta t \cr
                  -\eta       \sinh\eta t \cr
                   (\gamma/\eta)\sinh\eta t \cr }\right) .
  \cr}
\eqn\no
$$
The general real solution then reads
$$
  {\bfx}(t) = A {\bfh}_1 + B {\bfh}'_1(t) + C {\bff}_2(t)
              + D {\bff}_3(t) 
              + E {\bff}_4(t) + F {\bff}_5(t) \,,
\eqn\SOL
$$
with $A$, $B$, $C$, $D$, $E$, and $F$ some real constants.

To judge whether or 
not caustics occur in this system we have to examine
zero-modes, {\it i.e.}, 
classical solutions with vanishing boundary
conditions ${\bfx}(0)={\bfx}(T)={\bfzero}$. 
The initial condition put into (\SOL) gives
$$
    A = \omega (F-D), \quad
    C = {{\;\eta\;}\over\xi} \, E, \quad
    B = {{\eta\omega\beta}\over\gamma} \, E \ ,
\eqn\CoefCond
$$
and so the Jacobi field becomes
$D{\bfw_1}(t)+E{\bfw_2}(t)+
F{\bfw_3}(t)$ with 
$$ \eqalign{
    \bfw_1(t) & =
    \left(\matrix{ \omega     (\cos\xi t-1) \cr
                  -\xi        \sin\xi t     \cr
                  -(\gamma/\xi) \sin\xi t     \cr}\right) ,
  \cr
    \bfw_2(t) & =
    \left(\matrix{-(\omega/\eta)
( \sinh\eta t -\gamma^2 t/(\omega^2\eta)) \cr
                  -\cosh\eta t                         \cr
            (\gamma/\eta^2)(\cosh\eta t - 1)      \cr} \!
      \matrix{ + \cr + \cr + \cr}                    \!\!
          \matrix{ (\omega/\xi) 
      ( \sin \xi t  +\gamma^2 t/(\omega^2\xi) ) \cr
                   \cos\xi t                         \cr
       (\gamma/\xi^2) (\cos\xi t - 1)      \cr}\right) ,
  \cr
    \bfw_3(t) & =
    \left(\matrix{-\omega     (\cosh\eta t - 1) \cr
                  -\eta        \sinh\eta t     \cr
             (\gamma/\eta) \sinh\eta t     \cr}\right) .
  \cr
  }
\eqn\uvect
$$
The final condition
${\bfx}(T)={\bfzero}$ for a non-trivial choice of the
three remaining parameters 
$D$, $E$, and $F$ is fulfilled whenever the
matrix formed 
by the Jacobi fields becomes singular at $t=T$,
$$
\eqalign{
    0 & =
    \det\left( \bfw_1(T),
               \bfw_2(T),
               \bfw_3(T) \right) \cr
      & =
  \omega^3\beta\left[-{2\over\xi} (\cos\xi T-1) \sinh\eta T
               -{2\over\eta} \sin\xi T   (\cosh\eta T -1)
             +\beta T  \sin\xi T    \sinh\eta T \right]. \cr
}  
\eqn\vanishdet
$$
One sufficient condition for 
caustics is clearly seen to be
given by\note{%
Besides this,
there can be 
other caustics, {\it i.e.}, 
those which occur at $\xi\neq {\omega_n}$ 
for which the determinant (\vanishdet) vanishes.
}
$$
\xi = \xi(\omega) = \omega_n\ ,
\eqn\CAUSCOND
$$ 
with $\omega_n$ appearing in (\caustcondition),
which is indirectly a condition on the 
cyclotron frequency $\omega$.
In this case the coefficients in (\SOL) 
for the zero-mode solution read
$$
  A = -\omega D ,\quad 
  D = {\rm arbitrary},\quad 
  B = C = E = F = 0 \,.
\eqn\ZMD
$$
Since there is just one zero-mode, we have 
the partial caustics
situation $f=1$, $d=3$. The vectors $\bfw_i(t)$,
$i=1$, $2$, $3$, are related to 
the special choice of the Jacobi
fields ${\bfu}_i(t)$ in (\ucaust) 
to (\unorm) in the following way
$$
  \eqalign{
    {\bfu}_1(t) & =
    \phantom{{\;1\;}\over\sigma}M \;
    \bfw_1(t) ,
  \cr
    {\bfu}_2(t) & =
    {{\;1\;}\over\sigma} M
    \left[ {{\;\eta\;}\over\omega}\cdot
    {\sinh\eta T \over 1-\cosh\eta T}
           \,\bfw_2(t)
          +{{\;1\;}\over\omega}\,\bfw_3(t) \right] ,
  \cr
    {\bfu}_3(t) & =
    {{\;1\;}\over\sigma} M
    \left[ \bfw_2(t)
          +{{\;1\;}\over\eta}\cdot{\sinh\eta T
          -\eta\beta T \over {1-\cosh\eta T}}
           \,\bfw_3(t) \right] ,
  \cr}
\eqn\no
$$
with
$$
  \sigma := 
   2+ {{\eta\beta T \sinh\eta T} \over 1-\cosh\eta T}
  \,,\qquad
  M := \left(\matrix{0&0&1\cr 1&0&0\cr 0&1&0\cr}\right) \,.
\eqn\no
$$
The function ${\bfu}_1(t)$ is the 
zero-mode solution, and the caustics
surface is spanned by the two independent vectors
${\bfu}_2(T)$ and ${\bfu}_3(T)$, see Fig.\ 2.
On the other hand, 
the set of Jacobi fields ${\bfv}_i(t)$, $i=1,2,3$,
with the boundary conditions $v_i^j(0) = \delta_i^j$ 
and $\dot{v}^j_i(0) = 0$ 
introduced in (\vsol) can be expressed as follows
$$
  \eqalign{
    {\bfv}_1(t) & = {\bfh}_1 ,
  \cr
    {\bfv}_2(t) & =-{{\;\omega\;}\over\gamma}     {\bfh}'_1
              +{\xi \over\omega^2\beta} {\bff}_2(t)
              -{\eta\over\omega^2\beta} {\bff}_4(t) ,
  \cr
  {\bfv}_3(t) & = {\gamma\over\omega^2\beta\xi} {\bff}_2(t)
              +{\gamma\over\omega^2\beta\eta}{\bff}_4(t)
  \,.\cr
  }
\eqn\no
$$
\topinsert
\vskip 1.5cm
\let\picnaturalsize=N
\def\picsize{6cm}
\def\picfilename{f2.epsf}
\input epsf
\ifx\picnaturalsize N\epsfxsize \picsize\fi
\hskip 2.8cm\epsfbox{\picfilename}
\vskip 0.5cm
\abstract{%
{\bf Figure 2.}
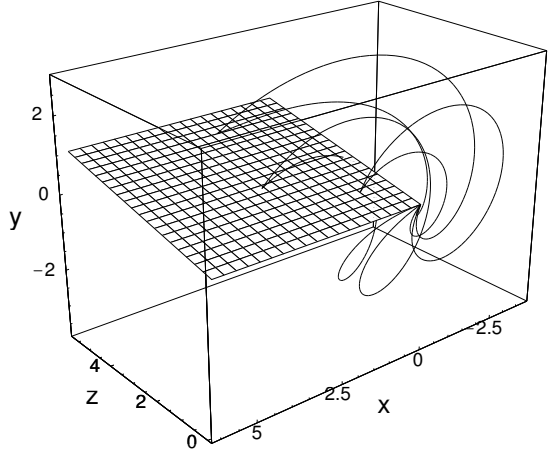
The behavior of the Jacobi fields and the caustics plane
for the parameter choice $\omega=1$, $\gamma = 0.2$.
For each of the three end points on the caustics plane we
can see two different paths starting at the origin, which
reflects the existence of the zero-mode ${\exbfu}_1(t)$.
}
\endinsert
\medskip

At this point it is worth mentioning 
the behavior of the Jacobi fields 
on caustics in the limit
$\alpha \rightarrow 0$, since in this limit 
the Lagrangian (\Lqa)
reduces to the Lagrangian (\L) 
of the previous subsection up to
the term $-q\phi$ 
linear in the coordinates, which does not
affect the behavior of the Jacobi fields.
One expects that the zero-mode 
solutions obtained in the previous
subsection for caustics $\omega = \omega_n$,
$$
  \eqalign{
    {\bfchi}_n{}^{\! (+)}(t)
     {\Big\vert}_{\omega=\omega_n} & =
    {1 \over 2}\sqrt{T \over 2}
    \left(\matrix{1-\cos\omega_n t \cr
                    \sin\omega_n t \cr}\right) ,
  \cr
    {\bfchi}_n{}^{\! (-)}(t)
     {\Big\vert}_{\omega=\omega_n} & =
    {1 \over 2}\sqrt{T \over 2}
    \left(\matrix{ -\sin\omega_n t \cr
                  1-\cos\omega_n t \cr}\right) ,
  \cr
  }
\eqn\ZERO
$$
are recovered as 
the $xy$-components of the first two of the Jacobi fields
in (\uvect), and, furthermore, that the twofold degeneracy
appears.
Indeed, since $\xi = \omega_n$ and
$$
  \left(\matrix{\gamma      \cr
                \eta        \cr
                \gamma/\eta \cr
                \omega      \cr}\right)
  \longrightarrow
  \left(\matrix{0           \cr
                0           \cr
                \omega      \cr
                \omega_n    \cr}\right)
  \qquad{\rm for}\qquad
  \alpha \rightarrow 0 \,,
\eqn\no
$$
we see that $-\bfw_1(t) / \omega$ goes over into the
upper vector and $-\bfw_2(t)$ 
into the lower vector of (\ZERO).
Whereas $\bfw_1(t)$ is already a zero-mode 
of the original system $\bfw_2(t)$ is not, 
which nonetheless turns into a zero-mode in the limit.
On the other hand, 
$\bfw_3(t) / \eta$ goes over into the function
$(0,0,t)^T$, 
which describes free motion of the $z$-component
in the direction of the caustics line.

When passed over to quantum mechanics, 
the transition amplitude for
the present partial caustics 
can also be obtained from the general formula (\finalkrl).  
However, we shall not record
it here as it becomes rather cumbersome due
to the structure of the matrix $Z(T)$ in (\zedd).

\vfill
\eject

\secno=4 \subsecno=0 \meqno=4 


\centerline{\bf 4. Conclusion}
\medskip

In this paper we studied caustics in 
quantum mechanics 
in multi-dimensions for Lagrangians which are
at most quadratic by 
extending our previous study in one-dimension
[\HMTT, \HMTTlet].  
Our main result is the kernel formula
(\finalkrl) which
gives the transition amplitude in $d$-dimensions.
As in one-dimension, we found that
transitions prohibited classically remain to be so even quantum 
mechanically.  
The complication that arises in the multi-dimensional
case is that we now have partial caustics, that is, the 
dimension of the caustics surface
where the classical trajectories (and hence the quantum 
amplitudes) concentrate
is not zero but smaller than $d$.  
More precisely, one looks at the number $f$ 
of zero modes of the operator defined by the
quadratic part of the Lagrangian, and if $f = d$ 
caustics take place maximally 
in all directions (full caustics),
if $f = 0$ there arises none (non-caustics), and otherwise 
we have partial caustics. For the full caustics case, the transition
kernel has been previously obtained in [\DM] based on a 
different formalism of the path-integral (\lq sum over all continuous
vector fields along a classical path vanishing at boundaries').
Here we have adopted a more intuitive and seemingly easier
method presented in [\Schulman], and thereby obtained the
kernel formula explicitly for generic caustics.
In the full
caustics case the kernel formula we obtained 
reduces to the one given in [\DM] as required, whereas 
in the non-caustics case it recovers 
the one (Gel'fand-Yaglom formula) which 
is familiar in semiclassical approximations.  
The crucial ingredient of our derivation is the unitarity
relation combined with the initial condition 
satisfied by the kernel, 
and to our amusement our method turns out to be
much simpler than the previous ones [\GY, \LS]
to get the well-known Van-Vleck 
formula for semiclassical approximations.

The two examples we presented illustrate how partial
caustics occur both classically and quantum mechanically.  
The first is the system of a 
charged particle under constant magnetic and electric fields
perpendicular to each other,   
which is found to be the case $d = 3$ and $f = 2$.  The 
caustics observed there is, however, 
not quite partial intrinsically,
because one can find an appropriate frame of coordinates
in which the partial caustics can be regarded as being a sum of 
$d=1$ non-caustics and $d=2$ full caustics.  
The second example is given by the system
of a charged particle under a certain 
(rather eccentric) electric field 
and provides the case $d = 3$ and $f = 1$.  In contrast
to the first example, this does not seem to admit 
a trivial decoupling into full and non-caustics beforehand 
and hence may be regarded as one which exhibits 
intrinsic partial caustics.

{}For further extension of the present
work, one obvious direction is to study quantum caustics 
in field theory, especially
in the context where the semiclassical approximation becomes
important.  This includes analyses of solitons/instantons
in, {\it e.g.,} the sine-Gordon theory, nonlinear sigma models,
and the Yang-Mills theory possibly coupled to various 
matter fields.  (In fact it was the question of the
role of intantons in the nonlinear sigma model [\Witten, \Jevicki]
which led us to the study of caustics originally [\HMTT].)
Another direction is to find applications
in condensed matter physics, where often situations
like the first example of Section 3 are considered.  To this
end we feel that the quantum feature of caustics should be
addressed in various physical aspects, 
such as in the spread of wave packets in the Gaussian slit
experiment [\HMTTlet]. 
We hope that
our present study serves as a basis for 
future investigations associated with caustics phenomena
including those mentioned here.

\bigskip

\noindent
{\bf Acknowledgement:}
We are grateful to S.~Tanimura for helpful 
discussions in the early stage of the work.

\ve
\bsk
\vfill\eject

  \vfill\eject\immediate\closeout\reffile
  \centerline{{\bf References}}\bigskip\frenchspacing%
  \input refs.tmp\vfill\eject\nonfrenchspacing

\bye